\documentclass[12pt]{spieman}  
\usepackage{amsmath,amsfonts,amssymb}
\usepackage{graphicx}
\usepackage{setspace}
\usepackage{tocloft}

\title{Inorbit Performance of the Hard X-ray Telescope (HXT) on board the Hitomi (ASTRO-H) satellite}

\author[a]{Hironori Matsumoto}
\author[b]{Hisamitsu Awaki}
\author[c]{Manabu Ishida}
\author[d]{Akihiro Furuzawa}
\author[e]{Shigeo Yamauchi}
\author[c]{Yoshitomo Maeda}
\author[f]{Ikuyuki Mitsuishi}
\author[g]{Yoshito Haba}
\author[f,h]{Takayuki Hayashi}
\author[c]{Ryo Iizuka}
\author[f]{Kazunori Ishibashi}
\author[i]{Masayuki Itoh}
\author[f]{Hideyo Kunieda}
\author[j]{Takuya Miyazawa}
\author[h]{Hideyuki Mori}
\author[h]{Takashi Okajima}
\author[k]{Satoshi Sugita}
\author[f]{Keisuke Tamura}
\author[f]{Yuzuru Tawara}

\affil[a,*]{Osaka University, Graduate School of Science, Department of Earth and Space Science, 1-1 Machikaneyama-cho, Toyonaka, Osaka, Japan, 560-0043}
\affil[b]{Ehime University, Graduate School of Science and Engineering, Matsuyama, Ehime, Japan, 790-8577}
\affil[c]{Japan Aerospace Exploration Agency, Institute of Space and Astronautical Science, Department of Space Astronomy and Astrophysics,
Sagamihara, Japan}
\affil[d]{Fujita Health University, School of Medicine, Toyoake, Japan}
\affil[e]{Nara Women’s University, Department of Physics, Faculty of Science, Nara, Japan}
\affil[f]{Nagoya University, Division of Particle and Astrophysical Science, Graduate School of Science, Nagoya, Japan}
\affil[g]{Aichi University of Education, Department of Science Education, Kariya, Japan}
\affil[h]{NASA’s Goddard Space Flight Center, X-ray Astrophysics Laboratory, Greenbelt, Maryland, United States}
\affil[i]{Kobe University, Graduate School of Human Development and Environment, Kobe, Japan}
\affil[j]{Okinawa Institute of Science and Technology Graduate University, Kunigami-gun, Okinawa, Japan}
\affil[k]{Tokyo Institute of Technology, School of Science, Department of Physics, Tokyo, Japan}

\cftpagenumbersoff{figure}
\cftpagenumbersoff{table} 
\begin{document} 
\maketitle

\begin{abstract}

  Hitomi (ASTRO-H) carries two Hard X-ray Telescopes (HXTs)
  that can focus X-rays up to 80~keV.  Combined with the
  Hard X-ray Imagers (HXIs) that detect the focused X-rays,
  imaging spectroscopy in the high-energy band from 5~keV to
  80~keV is made possible.  We studied characteristics of
  HXTs after the launch such as the encircled energy
  function (EEF) and the effective area using the data of a
  Crab observation. The half power diameters (HPDs) in the
  5--80~keV band evaluated from the EEFs are 1.59~arcmin for
  HXT-1 and 1.65~arcmin for HXT-2. Those are consistent with
  the HPDs measured with ground experiments when
  uncertainties are taken into account.  We can conclude
  that there is no significant change in the characteristics
  of the HXTs before and after the launch.  The off-axis angle of the
  aim point from the optical axis is evaluated to be less
  than 0.5~arcmin for both HXT-1 and HXT-2. The best-fit
  parameters for the Crab spectrum obtained with the HXT-HXI
  system are consistent with the canonical values.
  
\end{abstract}


{\noindent \footnotesize\textbf{*}Hironori Matsumoto,  \linkable{matumoto@ess.sci.osaka-u.ac.jp} }

\begin{spacing}{1}   

\section{Introduction}
\label{sect:intro}  

Hitomi (ASTRO-H)~\citenum{Tak18}, developed based on an
international collaboration led by ISAS/JAXA in Japan, was
launched on February 17, 2016. Hitomi carries two types of
X-ray telescopes; one is the Soft X-ray Telescope
(SXT)~\citenum{Soo11} that focuses X-rays below 10~keV, and
the other is the Hard X-ray Telescope (HXT)~\citenum{Awa14}
that can focus X-rays up to 80~keV. The HXT adopts a conical
approximation to the Wolter-I optics design
with a focal length of 12~m. Thin foils of aluminum
with a thickness of 0.2~mm and a height of 200~mm are used
as reflector substrates.  The radius of the innermost
reflector is 60~mm, and that of the outermost reflector is
225~mm.  The surface of the foils is covered with a
multilayer of platinum and carbon to reflect hard X-rays by
Bragg reflection. The total number of nesting shells is 213,
and the aperture is divided into three segments along the
azimuthal direction.  Since the HXT utilizes two-stage
reflection, the total number of reflectors is $213 \times 3
\times 2 = 1278$. There are 14 kinds of multilayers that are
applied to the HXT, and the details of the design is
described in Awaki et al. (2014)~\citenum{Awa14} and Tamura et
al. (2018)~\citenum{Tam18}.  There are two HXTs on board
Hitomi, and they are called HXT-1 and HXT-2.  The Hard X-ray
Imager (HXI)~\citenum{Nak18} is placed at the focal point of
each HXT, and HXI-1 and HXI-2 are combined with HXT-1 and
HXT-2, respectively.  Basic parameters of the HXT are summarized
in Table~\ref{tbl:HXTpara}.

\begin{table}
  \caption{Basic Parameters of the HXT.\label{tbl:HXTpara}}
  \begin{center}
    \begin{tabular}{cc} \hline \hline
      Number of telescopes & 2 (HXT-1 \& HXT-2) \\
      Focal length & 12~m \\
      Substrate & aluminum \\
      Substrate thickness & 0.2~mm \\
      Substrate height & 200~mm \\
      Coated multilayer & platinum and carbon \\
      Number of nesting shells & 213 \\
      Radius of innermost reflector & 60~mm\\
      Radius of outermost reflector & 225~mm\\
      Geometrical area & 968~cm$^2$/telescope \\ \hline
    \end{tabular}
  \end{center}
\end{table}

The HXTs were characterized with ground experiments mainly
done at the beam line BL20B2 of the synchrotron facility
SPring-8~\citenum{Mor18}. To analyze an X-ray spectrum of a
celestial object obtained with the HXT-HXI system, an
Ancillary Response File (ARF) that describes the specifics
of the response of the HXT such as an effective area for the object
and a Redistribution Matrix File (RMF) that 
describes the response of the HXI are needed.
The ARF is calculated by a raytrace program~\citenum{Yaq18}
based on the results of the ground experiments.
If the characteristics of the HXTs change after the launch,
the changes have to be incorporated into the ARF calculation.
For example, if the shape of the thin
foils change due to the release of gravitational stress in
orbit, the angular resolution and the effective area may change.
Although Hitomi experienced an attitude control anomaly and
was lost a month after the launch, the HXT-HXI
system was able to observe several objects in the first month.
In this paper, we studied the inorbit performance of the
HXTs using data from a Crab observation,
and compared the results with those obtained from the
ground experiments to see if any change occurred in the
characteristics of the HXTs after the launch.
Uncertainties are given at the $1~\sigma$ confidence
level unless otherwise stated in this work.

\section{Operation}

After the launch of Hitomi~\citenum{Tak18} on February 17,
2016, HXI-1~\citenum{Nak18} began to operate from March
12, 2016 when Hitomi targeted the high-mass X-ray binary
IGR~J$16318-4848$. The target was, however, outside the
field of view of the HXI-1. The HXI-2 was turned on on March
14 during the course of maneuver from IGR~J$16318-4848$ to
the neutron star RX~J$1856.5-3754$ which is one of the X-ray
dim isolated neutron stars with strong magnetic fields
\citenum{Tre00}.  The neutron star is known to
exhibit predominantly soft X-rays below 2~keV\citenum{Yon17}
and was observed for the calibration of the SXT\citenum{Soo11},
Soft X-ray Spectrometer (SXS)\citenum{Kel18}, and
Soft X-ray Imager (SXI)\citenum{Tan18} in the soft energy band.  The X-rays from
RX~J$1856.5-3754$ were too soft to be detected with the
HXT-HXI system. Hitomi observed the pulsar wind nebula
G~$21.5-0.9$ after RX~J$1856.5-3754$ on March 19, 2016.
Hard X-rays from this object were the first light for the
HXT-HXI system. Then Hitomi observed RX~J$1856.5-3754$ again
on March 22, 2016, and the Crab nebula was observed on March
25, 2016.
The exposure time of the Crab observation was about 8~ks.
The Crab nebula was the
2nd object that the HXT-HXI system detected hard X-ray
photons at its aim point.

Thus G~$21.5-0.9$ and the Crab nebula can be used to address
the inflight performance of the HXT. The hard X-ray emission
from G~$21.5-0.9$ is dominated by the pulsar wind nebula and
is spatially extended\citenum{Nyn14}.  Thus G~$21.5-0.9$ is
not an ideal target for characterizing the performance of
the HXT.  The hard X-ray emission of the Crab nebula is also
known to be spatially extended. However, the pulsation from
the Crab pulsar was successfully detected with the
HXI\citenum{Nak18}, while the pulsations from G~$21.5-0.9$
were not detected\citenum{Nyn14}. Using the pulse
information of the Crab pulsar, we can construct X-ray
images of the pulsar point source as described
below, and those images can be used to address the
performance of the HXT.  Thus we concentrate on the Crab
data in this paper.  The observation log of the Crab nebula
is summarized in Table~\ref{tbl:Crab}.

\begin{table}
  \caption{Observation log of the Crab nebula.\label{tbl:Crab}}
  \begin{center}
    \begin{tabular}{cccc} \hline
      Sequence Number & Observation start (UT) & Exposure (ks) & count rate (cts~s$^{-1}$) \\ \hline
      100044010 & 2016/3/25 12:37:19 & 8.01 & 397.9 (HXI-1), 403.0 (HXI-2) \\ \hline
    \end{tabular}
  \end{center}
\end{table}

\section{Encircled Energy Function\label{sec:EEF}}

\begin{figure}
\begin{center}
\includegraphics[width=0.6\textwidth]{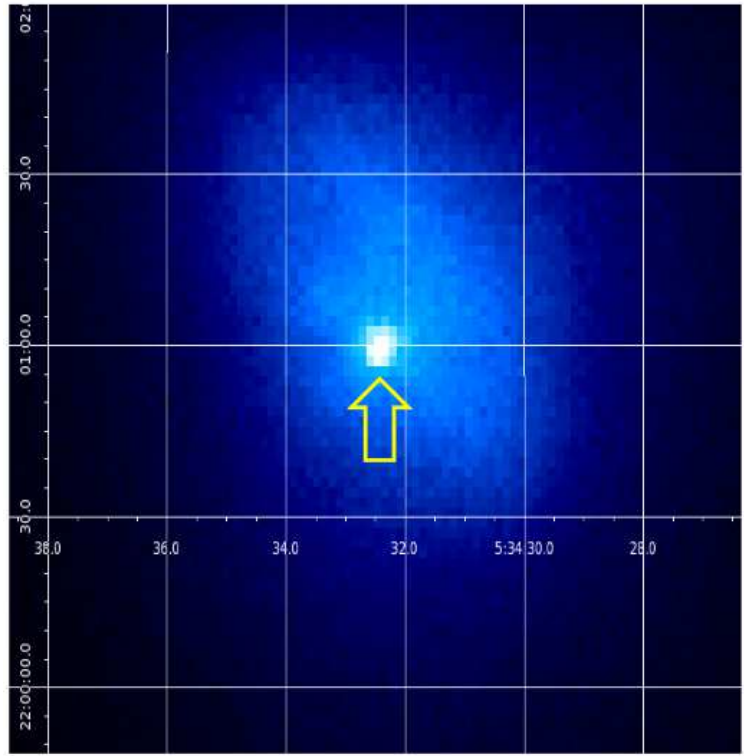}
\end{center}
\caption { 
  X-ray image of the Crab nebula in an energy band of 5 -- 80~keV
  obtained with the HXI-1.
  The position of the Crab pulsar is shown with an arrow.
\label{fig:crab_nebula_img}} 
\end{figure}

We used the cleaned event data of the Crab nebula with the
standard screening for the post-pipeline data
reduction\citenum{Ang16}. The sequence ID is 100044010, and
the processing script version of the data is 01.01.003.003.
The barycentric correction was applied where the target
position was the Crab pulsar position of $(\alpha,
\delta)_{\rm J2000} = (83.6332208,
22.0144614)$\citenum{Lyn93}.  The image of HXI-1 using all
cleaned data is shown in Fig.~\ref{fig:crab_nebula_img}.
The X-ray image consists of the Crab pulsar that is a point
source and the nebula around the pulsar that is spatially
extended.  In order to study the encircled energy function
(EEF) of HXTs, we need an X-ray image of a point source.
An X-ray image of the Crab pulsar excluding the
nebula emission was made as described below.

It is well known that the Crab pulsar exhibits X-ray
pulsations with a period of $\sim$33~ms.  The pulsations have a
double peak structure. The larger pulse is called P1 and the
smaller one is called P2\citenum{Kui01,Ge16}.  Since the HXI
has a time resolution of $25.6~\mu{\rm s}$\citenum{Nak18},
the pulsations were successfully detected by the HXI, and X-ray
images during the pulse P1 (phases 0.0--0.05 and
0.85--1.0) and during an off-pulse phase (phase 0.45--0.85; hereafter OFF1)
were obtained. If we subtract the OFF1 image from
the P1 image, we can obtain the X-ray image of the
Crab pulsar. However, since the count rate is large,
a dead time fraction has to be taken into account
in the subtraction process.
The method used for estimating the dead time fraction
is described in Appendix~\ref{sec:DTC}.

\begin{figure}
\begin{center}
\includegraphics[width=0.8\textwidth]{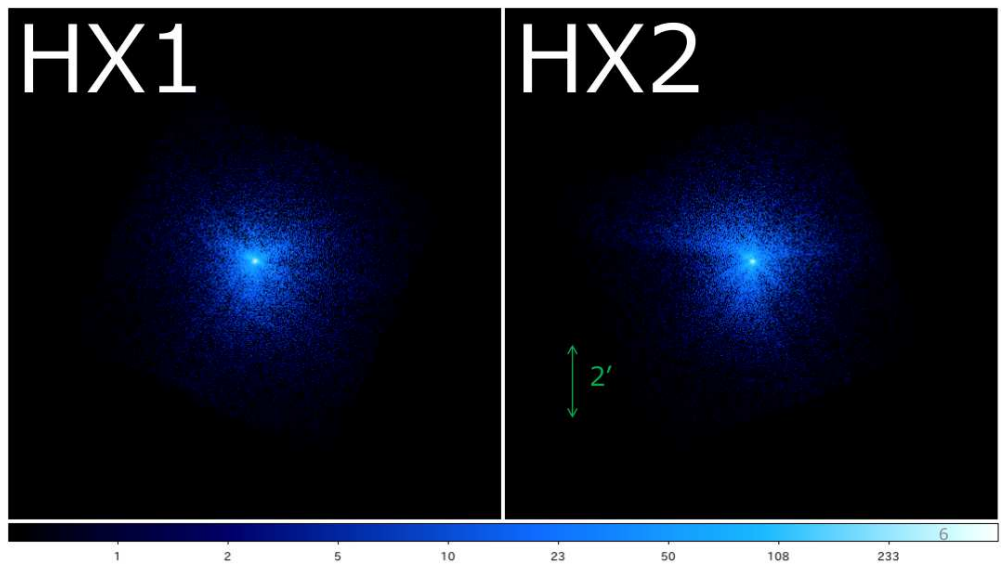}
\end{center}
\caption { 
  X-ray image of the Crab pulsar in the energy band of 5 -- 80~keV
  obtained with HXI-1 (left) and HXI-2 (right).
\label{fig:crab_pulsar_img}} 
\end{figure}

After obtaining the dead time fraction of each pulse phase,
X-ray images of the Crab pulsar was obtained by
applying a dead time correction.
For example, for the HXI-1, the exposure time of the OFF1 image was
$3.441 \times 10^3$~s, and the real exposure time was
$3.441 \times 10^3~{\rm s} \times (1 - 0.2210) = 2.681 \times 10^3~{\rm  s}$.
The exposure time of the P1 image was
$1.7205 \times 10^3$~s, and the real exposure time was
$1.7205 \times 10^3~{\rm s} \times (1 - 0.2571) = 1.278 \times 10^3~{\rm  s}$.
Then the Crab pulsar image of HXI-1 was obtained by
(P1 image) - (OFF1 image) $\times (1.278 \times 10^3~{\rm  s}/2.681 \times 10^3~{\rm s} )$.
The image of HXI-2 was obtained in the same manner.
The X-ray images in the 5 -- 80 keV~band thus obtained are shown in
Fig.~\ref{fig:crab_pulsar_img}.

\begin{figure}
\begin{center}
  \includegraphics[width=.48\textwidth]{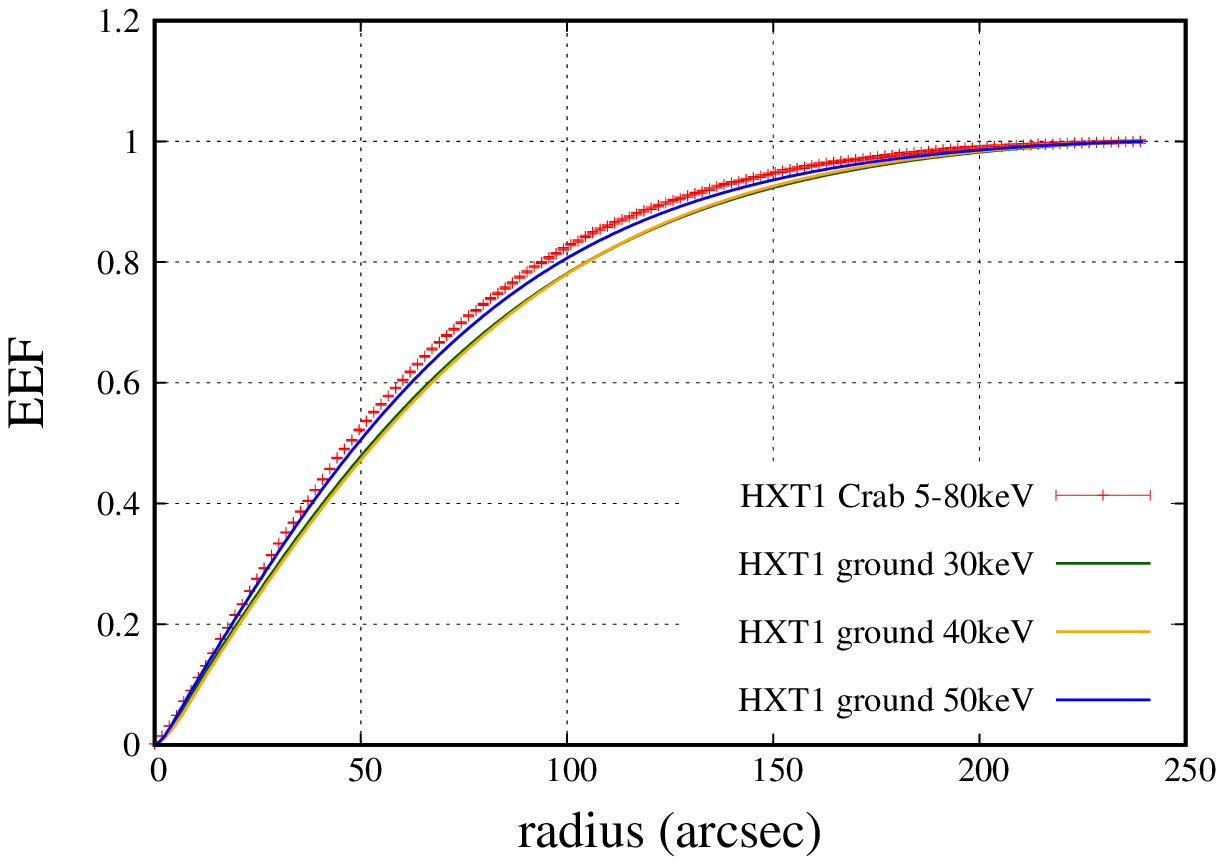}
  \hfill
  \includegraphics[width=.48\textwidth]{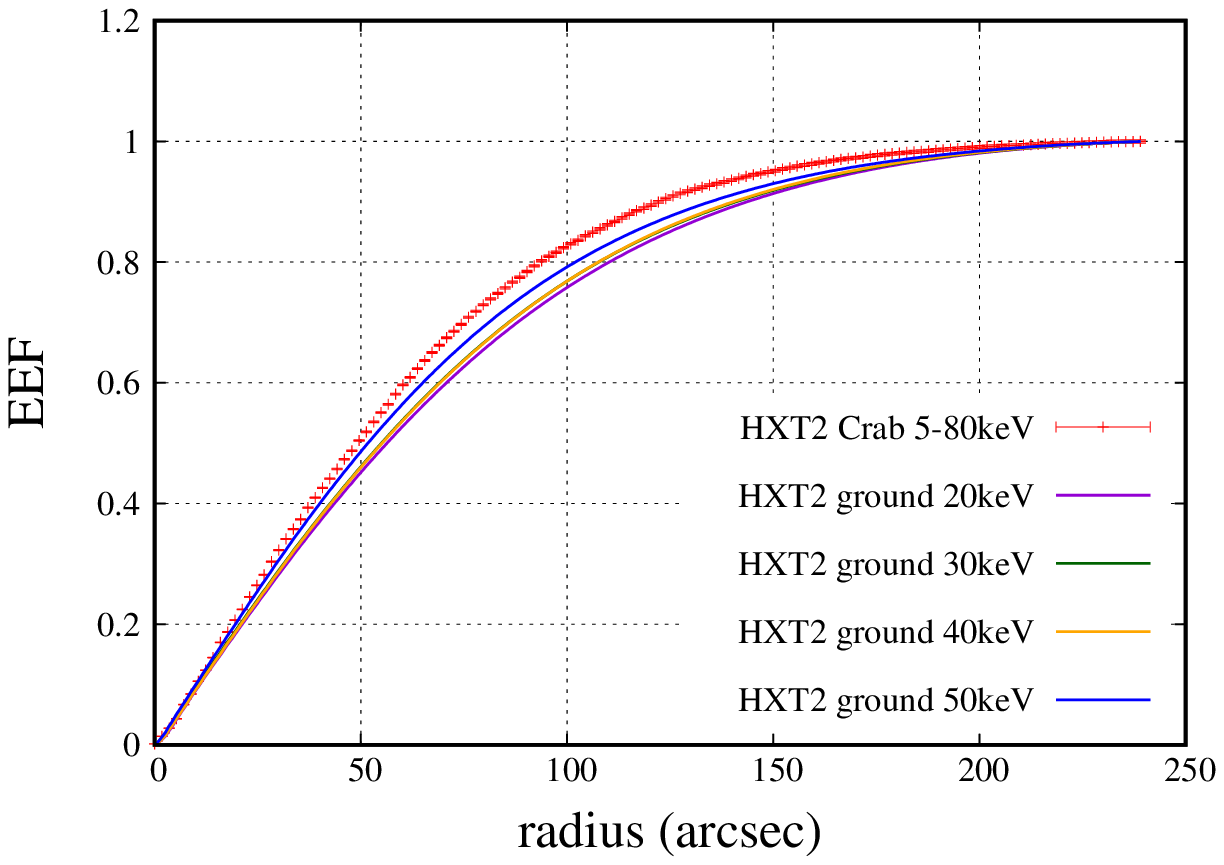}
\end{center}
\vspace{1em}
\caption { Encircled Energy Functions (EEFs) in the 5 -- 80~keV band of HXT-1
  (left) and HXT-2 (right) normalized at 4~arcmin are shown
  with red data points.  EEFs obtained with ground
  experiments at 20~keV (purple), 30~keV (green), 40~keV
  (orange) and 50~keV (blue) are also plotted.
  Error bars on the data are given at the $1~\sigma$ confidence level.
\label{fig:EEF_allE}} 
\end{figure}

\begin{figure}
\begin{center}
  \includegraphics[width=.48\textwidth]{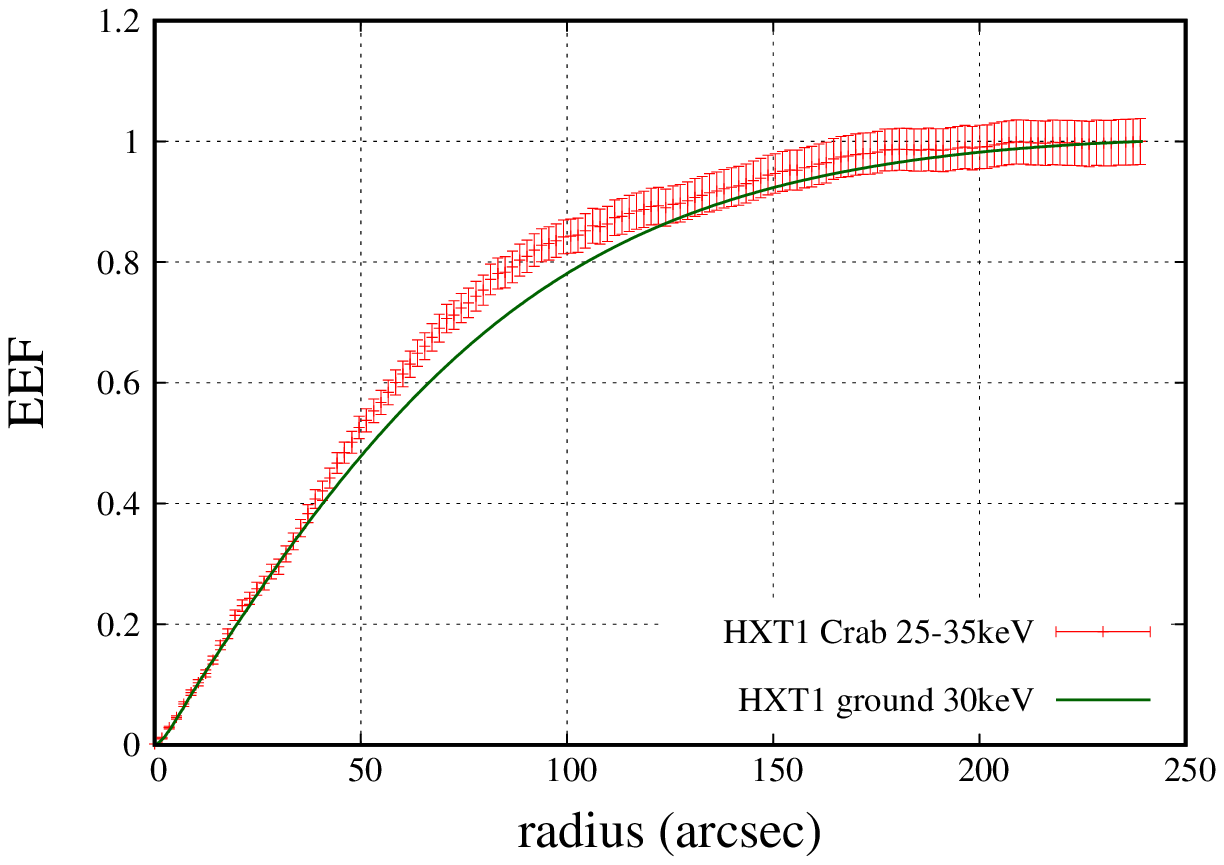}
  \hfill
  \includegraphics[width=.48\textwidth]{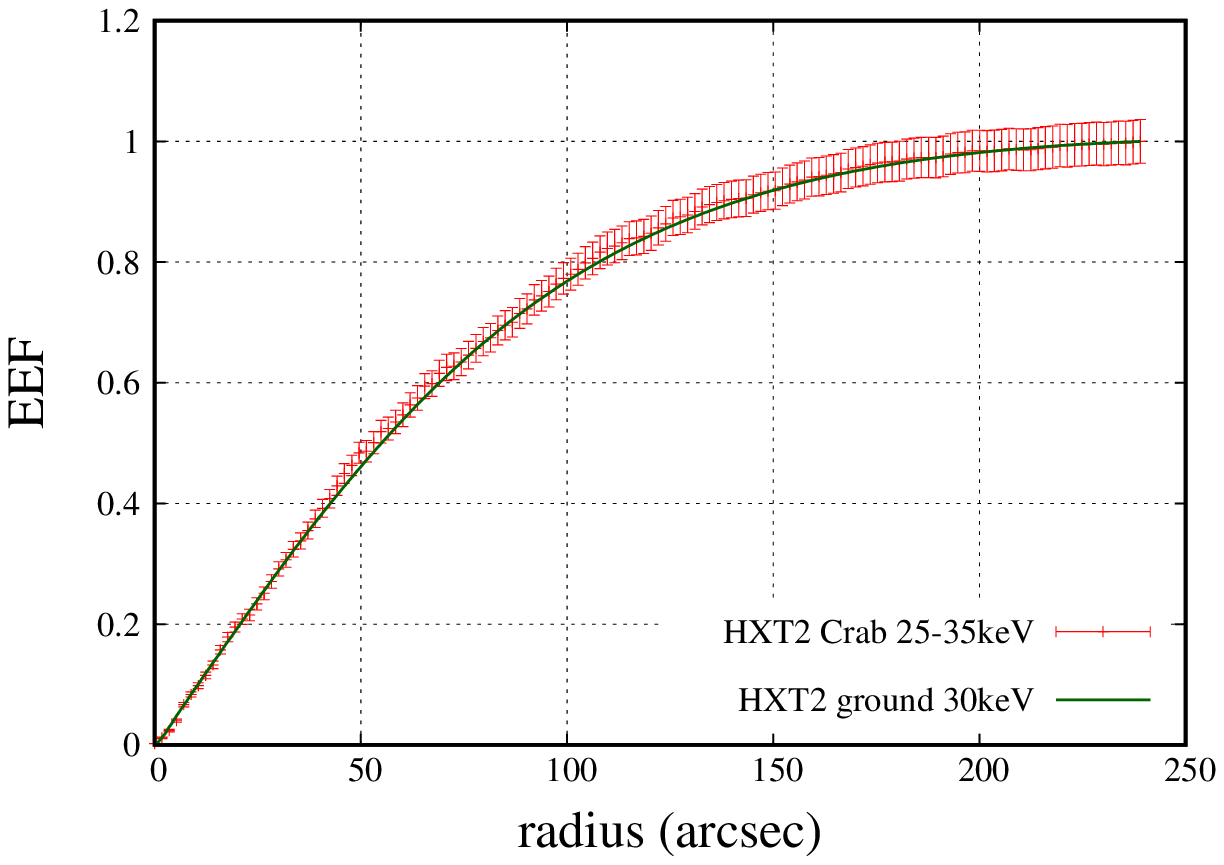}
\end{center}
\vspace{1em}
\caption { EEFs in the 25 -- 35~keV band of HXT-1
  (left) and HXT-2 (right) normalized at 4~arcmin are shown
  with red data points.  EEFs obtained with ground
  experiments at 30~keV (green) is also plotted.
  Error bars on the data are given at the $1~\sigma$ confidence level.
\label{fig:EEF_highE}} 
\end{figure}

The EEFs constructed from these images are shown in
Fig.~\ref{fig:EEF_allE}; in this case, the EEF is defined as
the ratio of the number of photons detected within a
circular region with a radius $r$ to those within a circle
with a radius of 4~arcmin. The center of the circular
regions coincides with the position of the Crab pulsar.
Since the field of view of HXI is 9~arcmin, the outermost
radius of the EEF is limited to 4~arcmin.  The average
number of photons per pixel in the field of view but outside
the circle of 4~arcmin radius was subtracted from the images
as background. The EEFs obtained by the ground experiments
done at SPring-8~\citenum{Mor18} are also plotted in
Fig.~\ref{fig:EEF_allE}; while the EEFs in Mori et
al. (2018)~\citenum{Mor18} are normalized at 6.2~arcmin, the
EEFs here are recalculated with the same procedure as that
used for the Crab image and are normalized at 4~arcmin.  The
half power diameter (HPD) is defined as the diameter where
the EEF equals to 0.5, and the HPDs are shown in
Table~\ref{tbl:HPD}.  Considering the uncertainty of the
dead time fraction, the uncertainty on the inorbit HPDs is
$\sim 0.1$~arcmin.  The HPDs from the ground experiments
also have an uncertainty of $\sim 0.1$~arcmin~\citenum{Mor18}.
The inorbit HPDs and those
from the ground experiments in Table~\ref{tbl:HPD} are
consistent with each other when the uncertainties are taken
into consideration.  Thus, the image quality in terms of the
HPD does not show a significant change between ground and in-orbit measurements.
Fig~\ref{fig:EEF_allE} suggests, however, that the inorbit
EEFs may be systematically narrower than those from the
ground experiments.  The release of the gravitational stress
may cause the slight contraction of the EEFs.  We
can conclude that the performance of HXTs did not change
significantly before and after the launch of Hitomi.  The
requirement on the HPD is 1.7~arcmin at
30~keV~\citenum{Awa14} using the EEF normalized at 6~arcmin,
and the requirement using the EEF normalized at 4~arcmin
corresponds to $\sim 1.6$~arcmin. To see inorbit HPDs at
30~keV, we examined the EEFs in the 25--35~keV band and they
are shown in Fig.~\ref{fig:EEF_highE}. The HPDs obtained
from those EEFs are also included in Table~\ref{tbl:HPD}.
Thus the HPD of HXT-1 in the 25--35~keV band satisfies the
requirement.
Although the HPD of HXT-2 is slightly larger than the
requirement, it is consistent with the
requirement within the uncertainties in the measurement.

\begin{table}
  \caption{Half Power Diameters\label{tbl:HPD}}
  \begin{center}
    \begin{tabular}{rcc} \hline
      &HXT-1& HXT-2 \\
      &(arcmin) & (arcmin) \\ \hline
      Inorbit 5--80~keV & 1.59 & 1.65 \\
      25 -- 35~keV & 1.59 & 1.77 \\
      Ground 20~keV & --- & 1.89 \\
      30~keV & 1.77 & 1.84 \\
      40~keV & 1.79 & 1.84 \\
      50~keV & 1.64 & 1.73 \\ \hline
    \end{tabular}
  \end{center}
\end{table}

\section{Off-axis angle of the aim point from the HXT optical axis\label{sec:aimpoint}}

The direction of the optical axis is defined as the
direction of the telescope at which the effective area is
maximized. It is required to observe an X-ray source with
various offset angles to determine the direction of the
optical axis. However, the Crab nebula was observed at the
aim point of the HXI, and we have no off-axis
observations. It is impossible to determine the direction of
the optical axis precisely. However, we can estimate the
off-axis angle of the aim point from the HXT optical axis by
comparing the spectrum of the Crab nebula with model
predictions calculated by the raytracing
program~\citenum{Yaq18} with different off-axis angles.

The cleaned data of the Crab nebula of the processing script
version 01.01.003.003 were used.
We extracted the spectrum of the
Crab nebula including all pulse phases from a circle with a
radius of 4~arcmin centered at the position of the Crab
pulsar.  The dead-time fraction was estimated using the pseudo
events. As is described in the appendix~\ref{sec:DTC}, this method gives
only a rough estimation of the dead-time fraction.
However, this estimation is enough for the
study in this section, since the
dead-time fraction is considered to have no energy
dependence and affects only the overall normalization of an
X-ray spectrum.
The background spectrum for each
sensor was obtained from blank sky observations.

\begin{figure}
  \begin{center}
\includegraphics[width=0.45\textwidth]{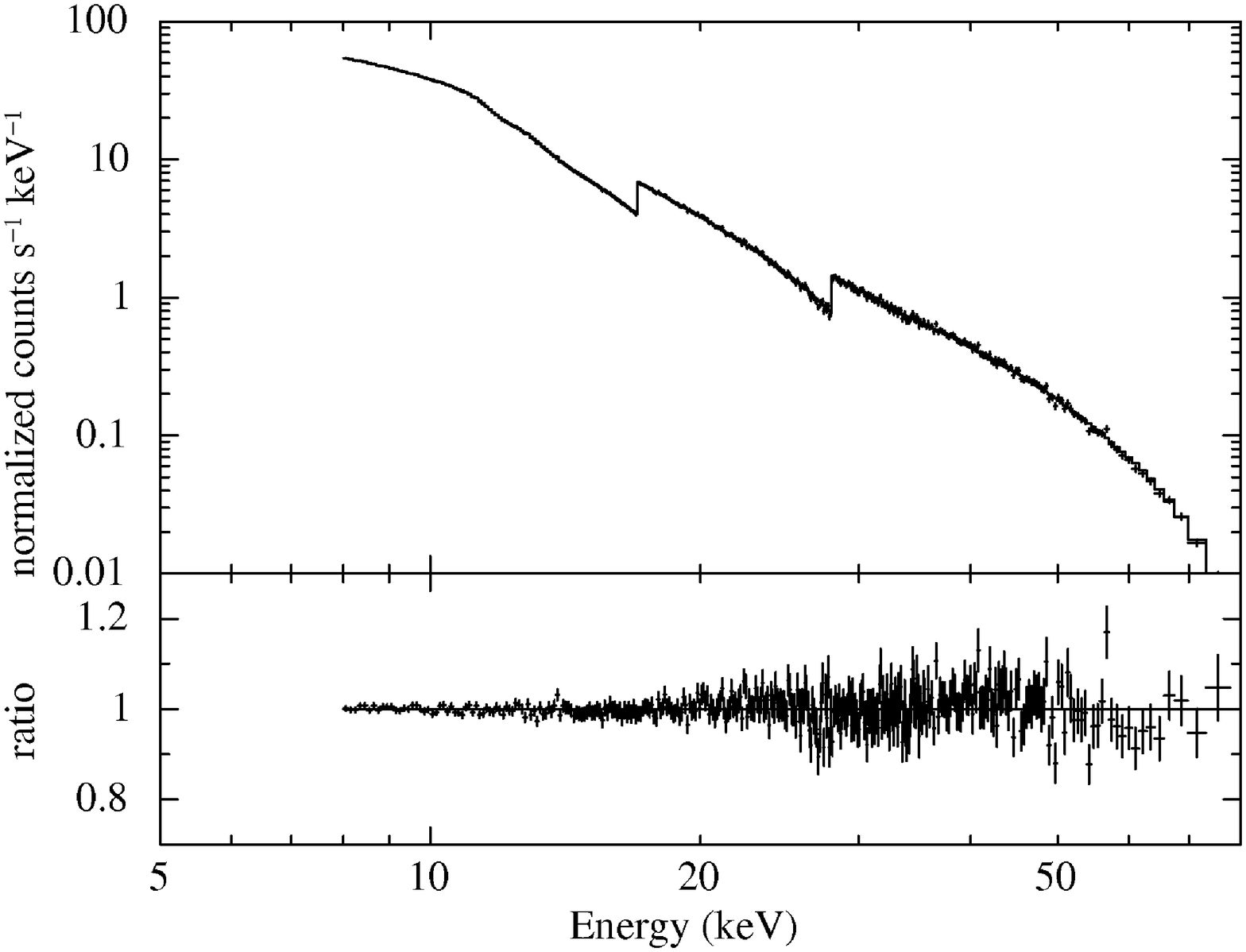}
\includegraphics[width=0.45\textwidth]{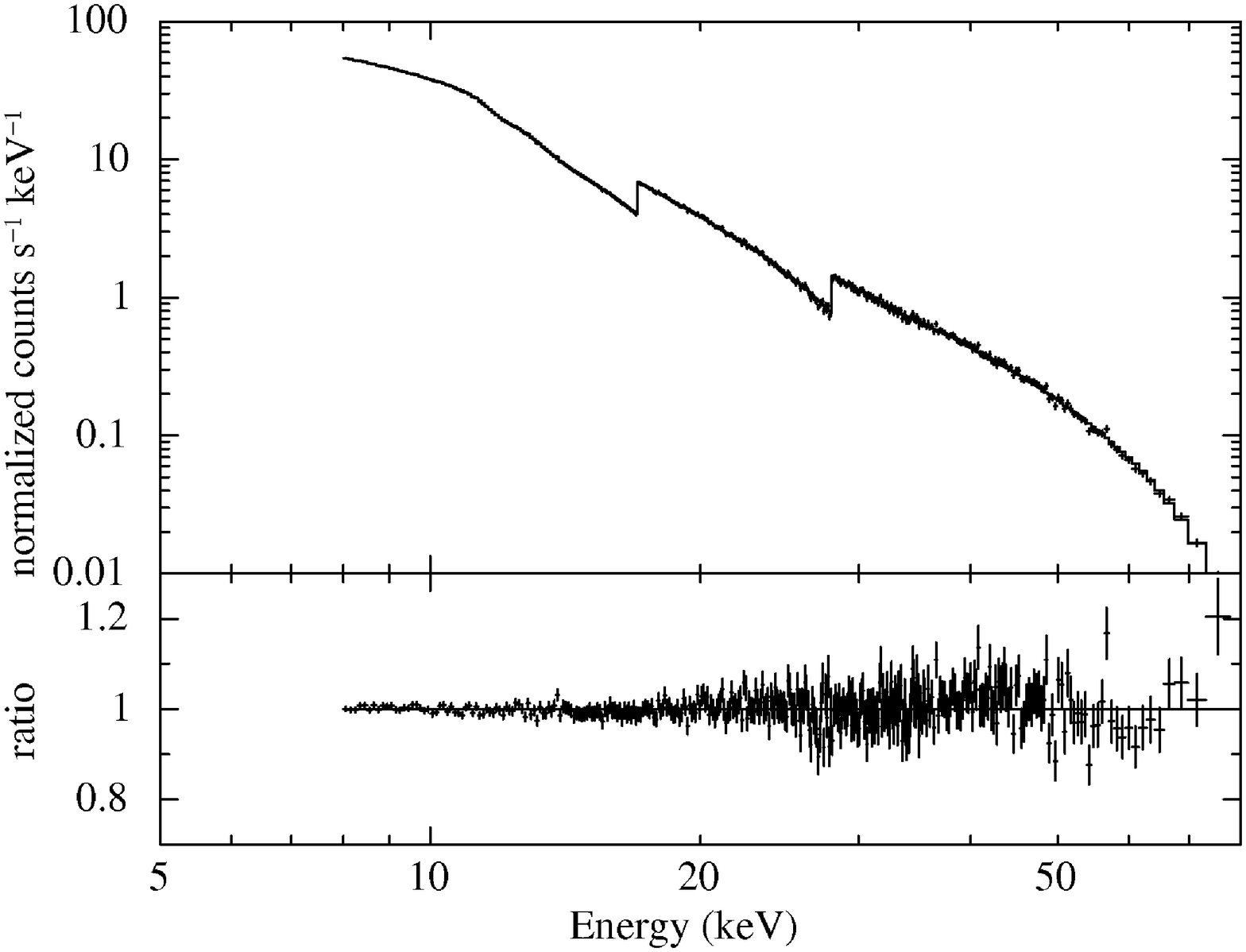}
\end{center}
\begin{center}
\includegraphics[width=0.45\textwidth]{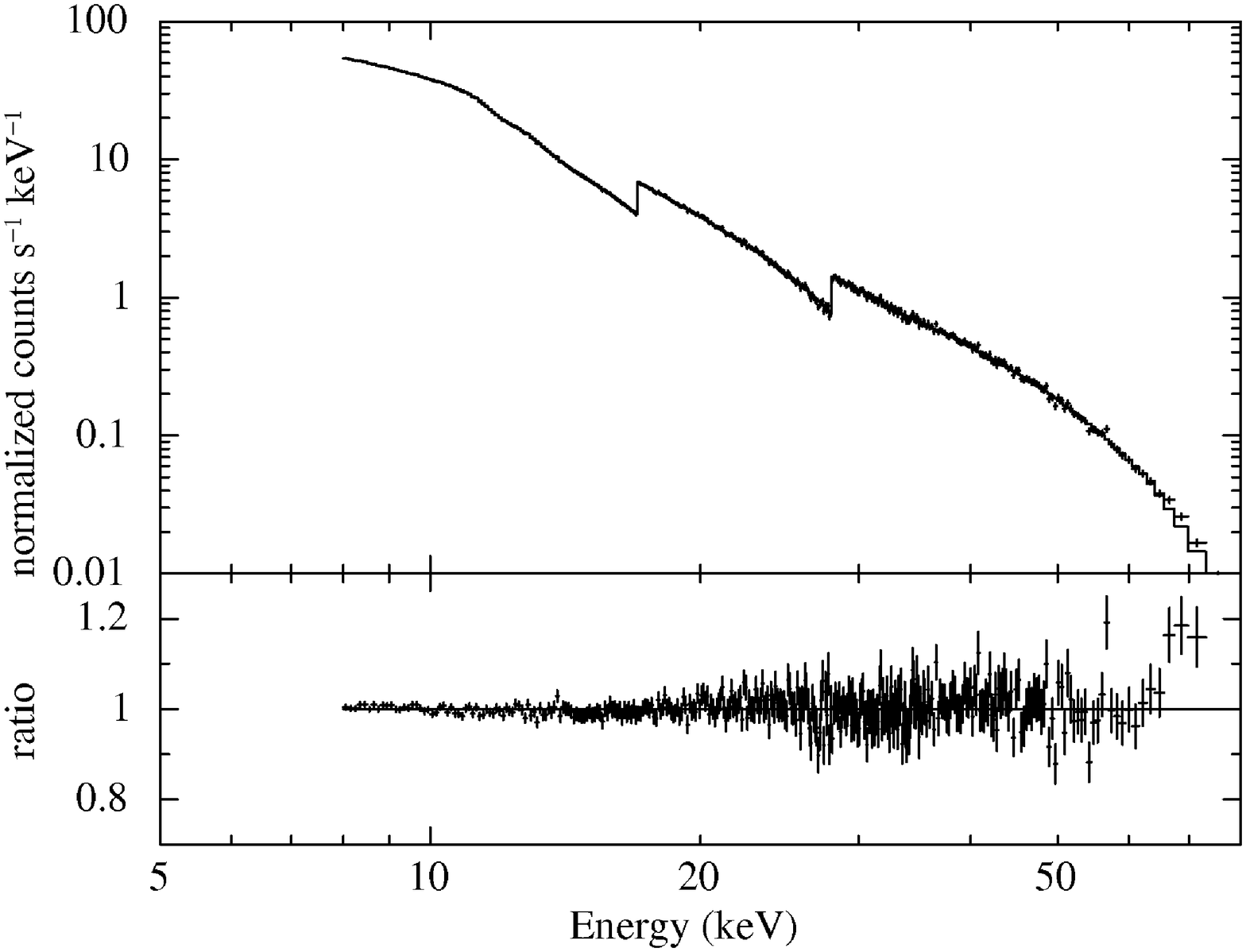}
\includegraphics[width=0.45\textwidth]{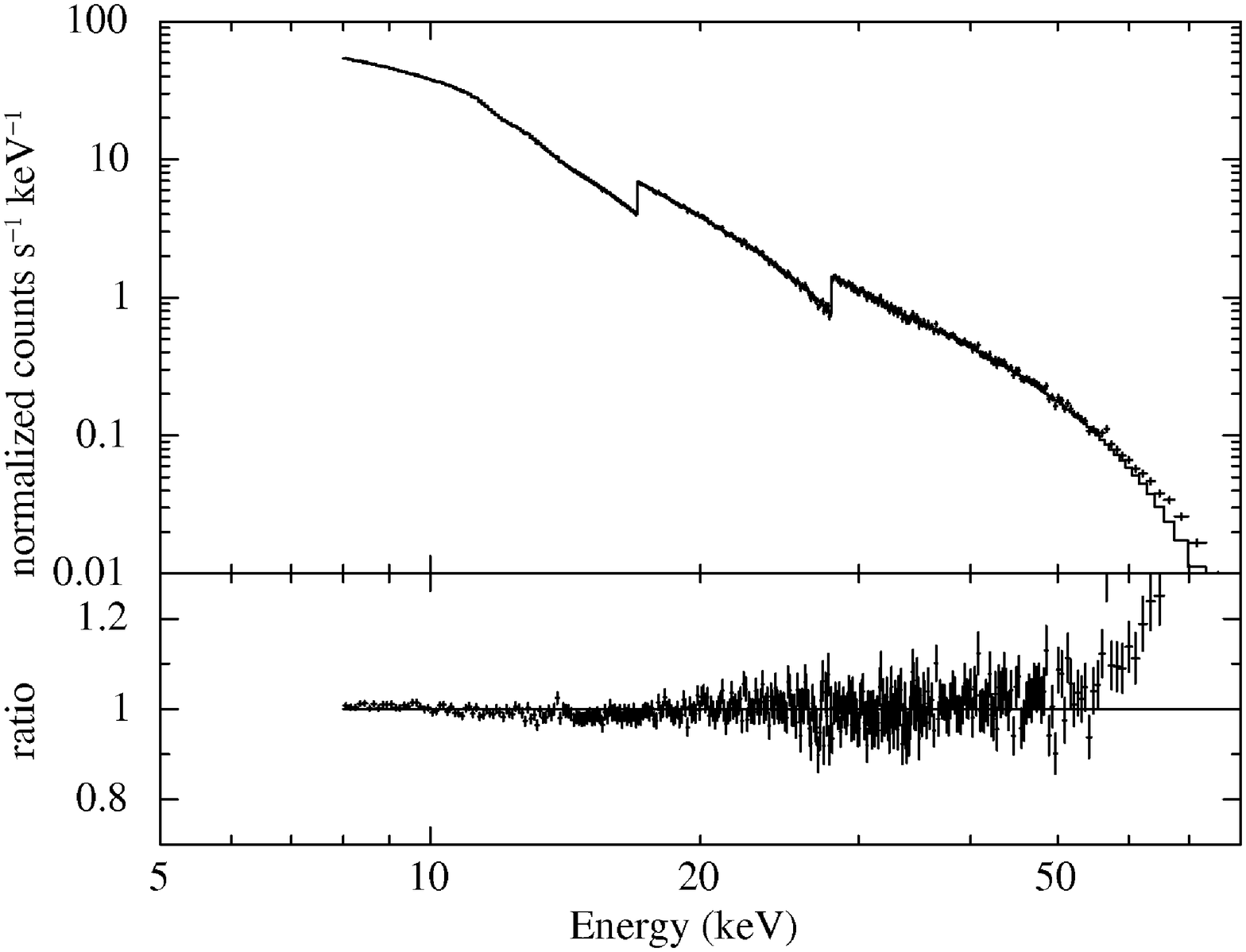}
\end{center}
\caption { HXT-1 spectrum of the Crab nebula in the 8 --
  80~keV band is fitted by a power-law model using ARFs
  assuming the off-axis angle of the Crab pulsar from the optical axis
  is 0.0~arcmin (top left), 0.5~arcmin (top
  right), 1.0~arcmin (bottom left), and 2.0~arcmin (bottom
  right). The lower panels show the ratio between the data
  and the best-fit model.
  Error bars on the data are given at the $1~\sigma$ confidence level.  
\label{fig:crab_spec_fit}} 
\end{figure}

A power-law model modified by the photoelectric absorption
of an equivalent hydrogen column density of $N_{\rm H} = 3
\times 10^{21}~{\rm cm}^{-2}$~\citenum{Kal05} was fitted to the spectrum
with various ARFs; the ARFs were created by assuming that
the off-axis angle of the Crab pulsar from the optical axis
is 0.0~arcmin, 0.5~arcmin, 1.0~arcmin, and 2.0~arcmin.  In
the ARF calculation, auxiliary transmission files
\verb+ah_hx[12]_auxtran_20140101v001.fits+ were used.
Though the raytrace program is based on the results of the
ground experiments, there are small differences (8~\% at
most) between the effective area measured at the ground
experiments and the area predicted by the raytrace
simulation.  The cause of the difference is not well
understood, and the auxiliary transmission files compensate
the differences by using arbitrary scaling factors.  The HXI-1 spectra with
the best-fit models are shown in
Fig.~\ref{fig:crab_spec_fit}. The best-fit parameters and
the $\chi^2$ values are listed in
Table~\ref{tbl:crab_spec_fit}. Note that the normalizations
in this Table are affected by the rough estimation of the
dead-time fraction. It is clear that an off-axis
angle of 1.0~arcmin or more cannot explain the spectrum at
the high energy side and the $\chi^2/d.o.f.$ value becomes
large as the off-axis angle increases.
This is because the vignetting function becomes narrower
as photon energy increases~\citenum{Awa14,Mor18}.
We can conclude that the off-axis angle of the aim point
from the optical axis is less than 0.5~arcmin.

\begin{table}
  \caption{Spectrum fit of the Crab nebula with ARFs assuming various off-axis angles.\label{tbl:crab_spec_fit}}
  \begin{center}
    \begin{tabular}{ccccc} \hline
      & \multicolumn{4}{c}{Off-axis angle for ARF of HXT-1} \\ 
      & on-axis & 0.5~arcmin & 1.0~arcmin & 2.0~arcmin \\ \hline
      Photon index & $2.114^{+0.004}_{-0.004}$ & $2.111^{+0.004}_{-0.004}$ & $2.093^{+0.004}_{-0.004}$ & $2.045$ \\
      Normalization & $10.39^{+0.09}_{-0.09}$ & $10.50^{+0.10}_{-0.09}$ & $10.52^{+0.10}_{-0.09}$ & $10.97$ \\
      Flux (2--10~keV) & $2.27 \times 10^{-8}$ & $2.30 \times 10^{-8}$ & $2.36 \times 10^{-8}$ & $2.65 \times 10^{-8}$\\
      $\chi^2 / d.o.f.$ & $354/360$ & $363/360$ & $397/360$ & $630/360$ \\ \hline
    \end{tabular} \\
    \vspace{1em}
    \begin{tabular}{ccccc} \hline
      & \multicolumn{4}{c}{Off-axis angle for ARF of HXT-2} \\ 
      & on-axis & 0.5~arcmin & 1.0~arcmin & 2.0~arcmin \\ \hline
      Photon index & $2.112^{+0.004}_{-0.004}$ & $2.111^{+0.003}_{-0.004}$ & $2.095^{+0.004}_{-0.004}$ & $2.053$ \\
      Normalization & $9.99^{+0.09}_{-0.09}$ & $10.12^{+0.09}_{-0.09}$ & $10.19^{+0.09}_{-0.09}$ & $10.87$ \\
      Flux (2--10~keV) & $2.18 \times 10^{-8}$ & $2.21 \times 10^{-8}$ & $2.28 \times 10^{-8}$ & $2.59 \times 10^{-8}$\\
      $\chi^2 / d.o.f.$ & $402/365$ & $416/365$ & $459/365$ & $825/365$ \\ \hline
    \end{tabular} 
  \end{center}
  \noindent
  Uncertainties are given at the 90~\% confidence level.\\
  Flux values are given in erg~s$^{-1}$~cm$^{-2}$.\\
  Normalizations are defined as the photon number flux at 1~keV.
\end{table}

\section{Inflight Performance and Raytrace\label{sec:inf_rayt}}

After establishing the off-axis angle of the aim point from
the optical axis is small, we can make the reasonable assumption
that the Crab pulsar was observed at the on-axis position.
Then the power-law model was fitted to the spectra of the Crab nebula
of HXI-1 and HXI-2 simultaneously. In this analysis, the cleaned data of
the Crab nebula of the processing script version
01.01.003.003 were reprocessed to be equivalent to those of
02.01.004.004.  The dead-time correction
described in the appendix~\ref{sec:DTC} was applied. The column density was
fixed to $N_{\rm H} = 3 \times 10^{21}~{\rm cm}^{-2}$.
The photon indices for both sensors were set to a common value,
while the normalizations for both sensors were varied separately.
The best-fit $\chi^2/d.o.f.$ is $738/679$.
The photon index is $2.122 \pm 0.003$, and the normalization,
which is defined as the photon number flux at 1~keV,
is $10.70 \pm 0.07$ for HXI-1 and $10.59 \pm 0.07$ for HXI-2,
where the uncertainties are given at the 90~\% confidence level.
These values are consistent with the ``canonical'' values
of the photon index $2.10 \pm 0.03$ and the normalization
$9.7 \pm 1.0$ in Toor and Seward (1984)~\citenum{Too74}.
The difference of the normalizations of HXI-1 and HXI-2
is $\sim 1$~\%.
The unabsorbed energy flux in the 3--50~keV band
calculated from the normalization is $3.59 \times 10^{-8}~{\rm erg~cm^{-2}~s^{-1}}$ for HXI-1 and
$3.55 \times 10^{-8}~{\rm erg~cm^{-2}~s^{-1}}$ for HXI-2.
These values are $\sim 5\%$ larger than the flux of
$3.37 \times 10^{-8}~{\rm erg~cm^{-2}~s^{-1}}$ measured with NuSTAR~\citenum{Mad17}.

\begin{figure}
\begin{center}
\includegraphics[width=0.8\textwidth]{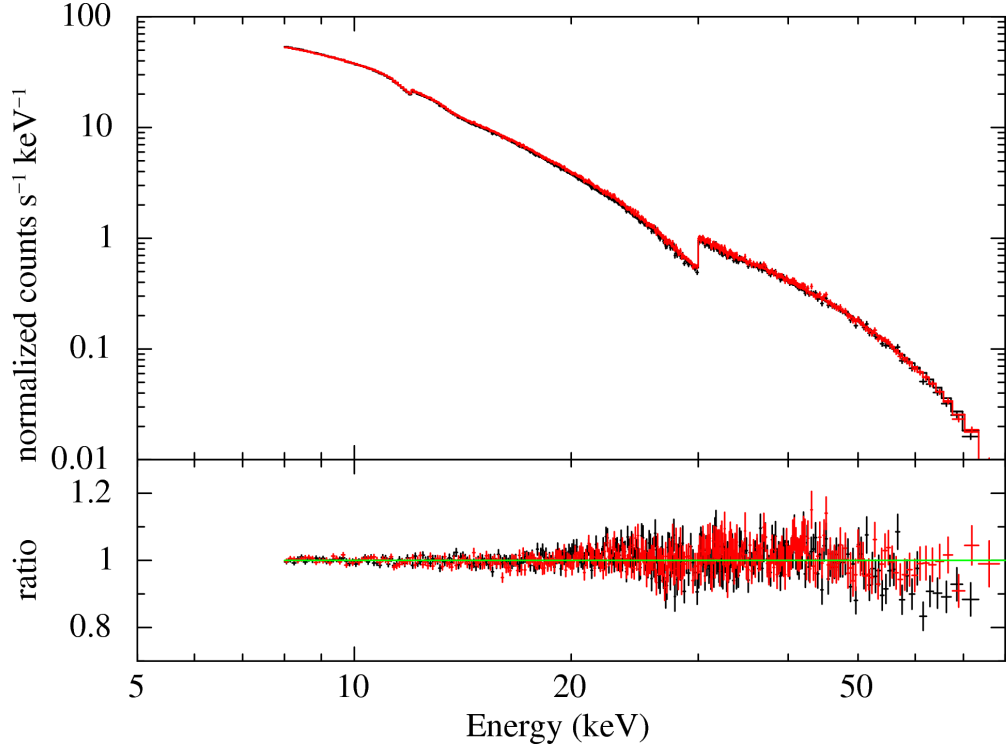}
\end{center}
\caption { Spectra of the Crab nebula in the 8 --
  80~keV band of HXI-1 (black) and HXI-2 (red) together
  with the best-fit power-law model using the on-axis ARFs.
  The lower panels show the ratio between the data
  and the best-fit model.
  Error bars on the data are given at the $1~\sigma$ confidence level.  
\label{fig:crab_spec_fit_both}} 
\end{figure}

\begin{figure}
\begin{center}
  \includegraphics[width=0.45\textwidth]{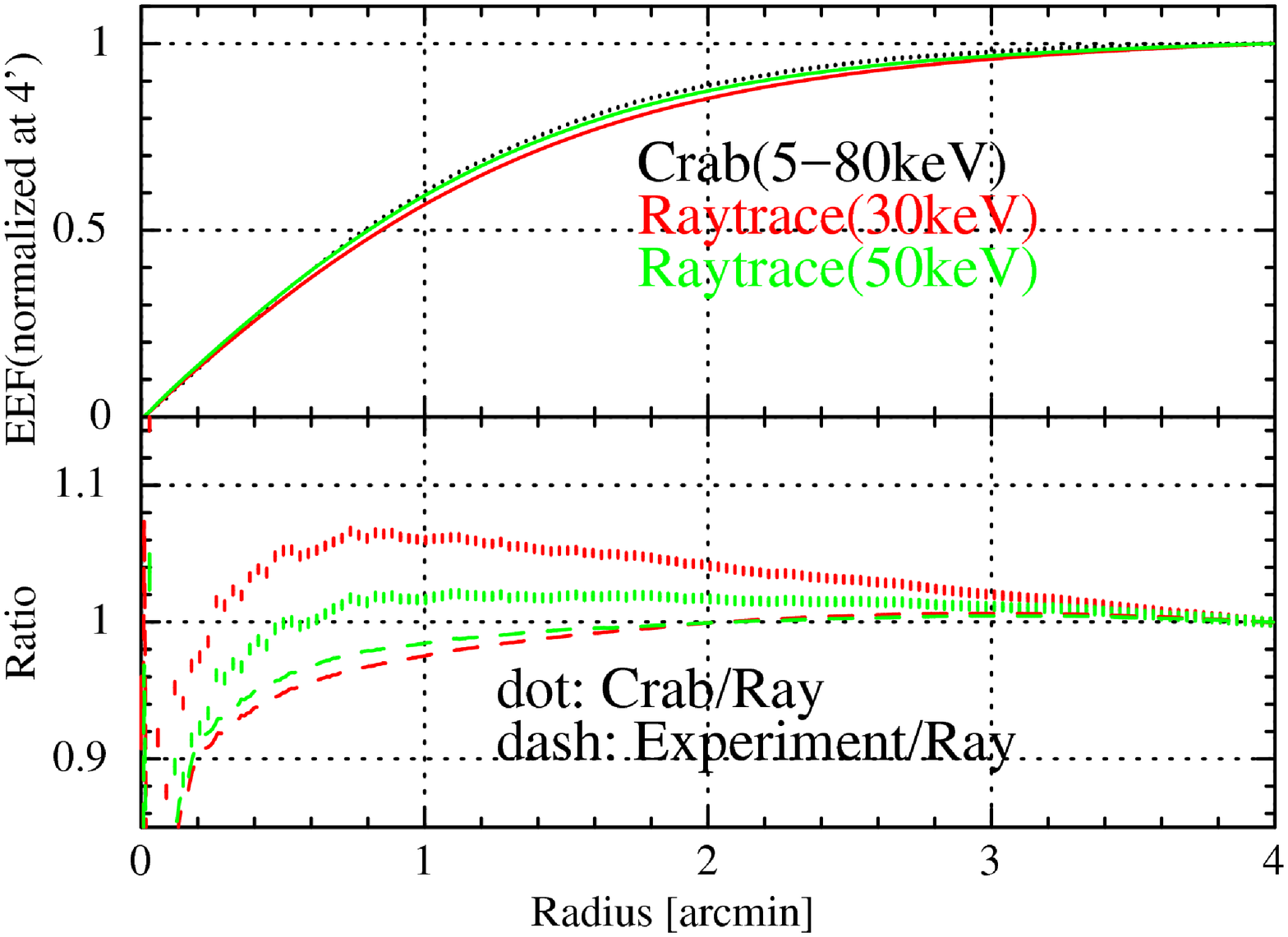}
  \includegraphics[width=0.45\textwidth]{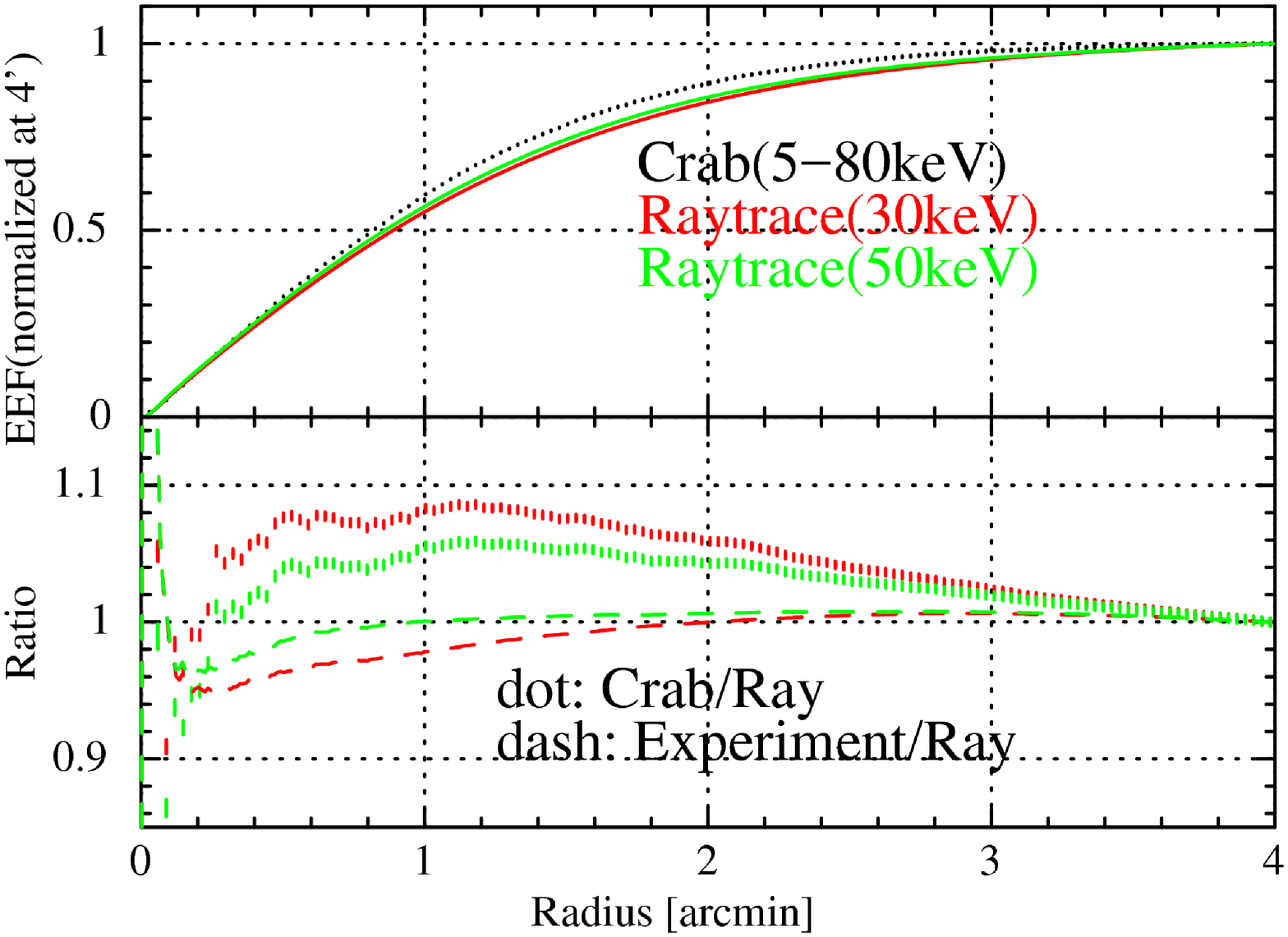}
\end{center}
\caption { EEF of the Crab pulsar in the 5--80~keV band (black) is compared with
  those predicted by the raytrace program at 30~keV (red)
  and 50~keV (green) in the top panel: HXT-1 (left) and HXT-2 (right).
  Ratios between them are plotted with the dotted line in the bottom panel,
  while ratios of the EEFs obtained by the ground experiment
  to the raytrace EEFs are shown as the dashed line.
  Error bars on the data are given at the $1~\sigma$ confidence level.  
\label{fig:eef_crab_rayt}} 
\end{figure}

Fig.~\ref{fig:eef_crab_rayt} shows
the EEF of the Crab pulsar in the 5--80~keV band generated
in section~\ref{sec:EEF} together with those predicted by the
raytracing program.
The bottom panel shows the ratio between them as dotted lines.
The ratios between the EEFs from the ground experiments
and from the raytrace calculations are also plotted as dashed lines.  The
deviation between the raytrace EEFs and the Crab EEF is less
than 10~\% except for the central region within a radius
0.2~arcmin. We should note that if we extract an HXI
spectrum from a circular region with a small radius,
the effective area for the spectrum calculated using the raytrace
program has a systematic uncertainty.

\section{Summary\label{sec:summary}}

The X-ray image of the Crab pulsar point source
for the HXT-HXI system was obtained by subtracting the Crab
image during the off-pulse (OFF1) phase from that during the
on-pulse (P1) phase. In the subtraction process, the dead
time fraction was taken into account.  The EEF normalized
at a radius of 4~arcmin was constructed from the image.
The HPD was estimated from the EEF,
and the HPDs in the 5--80~keV band are 1.59~arcmin for HXT-1
and 1.65~arcmin for HXT-2.  These HPDs are consistent with those obtained from
the ground experiments within the uncertainty, and this
suggests that there is no significant change in
the characteristics of HXTs before and after
the launch of Hitomi.
We estimate that the
off-axis angle of the aim point from the direction of the
optical axis is less than 0.5~arcmin for both HXT-1 and
HXT-2 as determined by the model fitting of the spectrum of the Crab
nebula using ARFs assuming various off-axis angles.  The
best-fit parameters for the Crab spectrum are consistent
with the canonical values of Toor \& Seward\cite{Too74}. The
deviation between the inorbit EEF and those calculated by
the raytrace program is less than 10~\% except for the
region with a radius smaller than 0.2~arcmin.

\appendix

\section{Evaluation of the Dead Time Fractions\label{sec:DTC}}

The HXI has pseudo events that are randomly distributed
with a mean frequency of 2~Hz. The number of pseudo
events gives an estimate of the exposure time after the dead
time loss. Since the total exposure time is 8~ks and then
the exposure time of each pulse phase is only a few ks,
however, the Poisson fluctuation of the pseudo events
dominates the uncertainty in the estimation of the dead time
fraction of each pulse phase.  For example, the dead time
fraction of HXI-1 during the pulse P1 phase was estimated to
be 24.4~\% by using the pseudo events, while that during the
off-pulse phase was to be 24.5~\%. Of course the former
should be larger than the latter, and this means that the
uncertainty in the dead time fraction estimated by the pseudo
events was not small. Then we used another method described
below to estimate the dead time fractions.

First we estimated a typical dead time of each event.  We
investigated the distribution of \verb+LIVETIME+ tagged to each
event; the livetime of the event is the time interval
between the end time of the processing of the previous
trigger and the trigger of this event. The distribution
should be proportional to $\exp (-nt)$, where $t$ is the
livetime and $n$ is the ``true'' count rate which is the
rate that would be recorded if there were no dead
time\citenum{Kno99}.  The distribution of the live time of
all events suggests that $n$ is 726~c~s$^{-1}$ for HXI-1.  The
uncertainty of $n$ at the 1~$\sigma$ confidence level is $\sim\ 0.1~\%$.
If a count rate $m$
recorded after suffering the dead time loss is obtained, we
can calculate the dead time per event $\tau$ by $\tau =
(1/m) - (1/n)$\citenum{Kno99}. All events are classified on
board as a category H, M, or L, where the category H is a
normal event, and the categories M and L are the events that
coincide with a signal from the active shield counter. The
number of events in each category can be found in the HK
file as \verb+HXI[12]_USER_EVNT_SEL_CNT_[HML]+.  Using the
HK information, the averaged total count rate during the
Crab observation recorded with HXI-1 was estimated to be $m
= 537$~c~s$^{-1}$.  The uncertainty of $m$ at the 1~$\sigma$ confidence level
is $\sim\ 0.1~\%$.
Then the typical dead time $\tau = (1/537) - (1/726) = 3.678 \times 10^{-4}~{\rm s~c}^{-1}$.

Next, we evaluated the recorded count rate of each pulse
phase. However, the time resolution of the HK information is
not sufficient for this purpose. Then we evaluated count rates
of the three categories separately and added them.  The
count rate of the category H can be estimated by just
counting the event number in unfiltered event files.
However, most of the events of the categories M and L are
not included in the event files, since their priority is low
and most of them are not kept in the onboard data recorder
to save the capacity of the recorder. As for the count rates
of the categories M and L, we assumed that they consist of
two parts; one is a background constant component which can
be measured during the Earth occultation, and the other part
comes from events accidentally coincide with the shield
events. We assumed that the latter is proportional to the
count rate of the category H after subtracting its constant
component during the Earth occultation. In summary, we assumed
\begin{equation}
  M + L = a \times (H - H_0) + (M_0 + L_0),
  \label{eq:1}
\end{equation}
where $a$ is a constant, $H$, $M$, and $L$ denote the count
rate of each category respectively, and the subscript $0$
means that those are rates during the Earth
occultation. $H_0 = 18.47$~c~s$^{-1}$ and $M_0 + L_0 = 15.19$~c~s$^{-1}$
were obtained for HXI-1 from the HK
information. The averaged $H$ and $M + L$ during the Crab
observation for HXI-1 were also obtained from the HK
information, and $H = 536.3$~c~s$^{-1}$ and $M + L = 36.49$~c~s$^{-1}$.
From these values, the constant $a$ for
HXI-1 was estimated to be $0.0409$.  Then $M + L$ of each
pulse phase was estimated from $H$ in each pulse phase,
and we added them to obtain $H + M + L$.

Though a measure of the dead time fraction can be obtained
by $DT' = (H + M + L) \times \tau$, one more correction has
to be taken into account. Since \verb+LIVETIME+ is the time
interval between H events, it does not reflect the fact that
some of the H events accidentally coincide with the shield
events and are classified as M or L. The real live time of
these accidental M or L events is smaller than their \verb+LIVETIME+.
The accidental event rate corresponds to $a \times (H - H_0)$ in
equation~(\ref{eq:1}). The real dead time fraction was
estimated by $DT = 1 - (1 - DT') \times (1 - f)$, where $f$
is defined as the fraction of the accidental events and is
calculated by $f = (a \times (H - H_0))/(H + M + L)$.  The
final results are shown in Table~\ref{tbl:DTF}.  Typical
uncertainties on $DT$ at the 1$\sigma$ confidence level is
$\sim\ 2~\%$. The same procedure was applied for the HXI-2
data, and the results are also listed in Table~\ref{tbl:DTF}.
See the paper on the inflight performance of the HXI\citenum{Hag18}
for more detailed information.

\begin{table}
  \caption{Dead time fraction of different pulse phases.\label{tbl:DTF}}
  \begin{center}
    \begin{tabular}{lcccc} \hline \hline
      Phase & $H+M+L$ (c~s$^{-1}$) & $DT'$ & $f$ & $DT$ \\ \hline
      \multicolumn{5}{c}{HXI-1 ($\tau = 3.678 \times 10^{-4}$~s~c$^{-1}$)} \\
      0.0--0.05, 0.85 -- 1.0 (P1) & 619.75 & 0.2279 & 0.0372 & 0.2571 \\
      0.05--0.2 (OFF2) & 547.99 & 0.2015 & 0.0370 & 0.2315 \\
      0.2--0.45 (P2) & 612.87 & 0.2254 & 0.0372 & 0.2547 \\
      0.45 -- 0.85 (OFF1) & 518.62 & 0.1908 & 0.0368 & 0.2210 \\
      average & 566.75 & 0.2085 & 0.0370 & 0.2382 \\
      \multicolumn{5}{c}{HXI-2 ($\tau = 3.744 \times 10^{-4}$~s~c$^{-1}$)} \\
      0.0--0.05, 0.85 -- 1.0 (P1) & 662.78 & 0.2438 & 0.0332 & 0.2692 \\
      0.05--0.2 (OFF2) & 588.25 & 0.2164 & 0.0329 & 0.2425 \\
      0.2--0.45 (P2) & 654.21 & 0.2406 & 0.0332 & 0.2661 \\
      0.45 -- 0.85 (OFF1) & 557.68 & 0.2051 & 0.0328 & 0.2316 \\ 
      average & 607.42 & 0.2234 & 0.0330 & 0.2494 \\ \hline
    \end{tabular}
  \end{center}
\end{table}

\acknowledgments

We appreciate all the people who contributed to the Hitomi project.
We are especially grateful to people at Tamagawa Engineering, LTD.,
especially Naoki Ishida, Hiroyuki Furuta, Akio Suzuki, and Yoshihiro Yamamoto,
for their invaluable contribution to the production of the HXTs.
We also thank Kaori Kamimura, Ayako Koduka, Yuriko Minoura,
Yumi Mori, Keiko Negishi, Megumi Sasaki, Yasuyo Takasaki, and Tamae Yamagishi
for their support to the HXT production.
We acknowledge the support from the JSPS/MEXT KAKENHI program.
Their grant numbers are 
15H02070, 15K13464, 24340039, 22340046, 23000004, and 19104003.
Also we were supported from the JST-SENTAN Program 12103566.
This research used many results of the experiments
performed at the BL20B2 of SPring-8
with the approval of the Japan Synchrotron Radiation
Research Institute (JASRI) as PU2009A0088,
``Development of a System for Characterization of Next-generation X-ray Telescopes for Future X-ray Astrophysics''.


\bibliography{HXT_ref}   
\bibliographystyle{spiejour}   


\vspace{2ex}\noindent\textbf{Hironori Matsumoto} is a professor
at Osaka University. He received his BS, MS, and PhD degrees in physics
from Kyoto University in 1993, 1995, 1998, respectively.
His research filed is X-ray astronomy including
developing instruments.
He is a member of SPIE.


\end{spacing}
\end{document}